\documentclass[review]{elsarticle}

\usepackage{hyperref}
\journal{Journal of \LaTeX\ Templates}

\usepackage{algorithm}
\usepackage{algorithmic}

\usepackage{amsmath}
\usepackage{multirow}

\usepackage{subfigure}
\usepackage{graphicx}
\usepackage{epsfig}
\usepackage{epstopdf}
\usepackage{amssymb}
\usepackage{bm}
\usepackage{booktabs}
\usepackage{verbatim}
\usepackage{array}

\begin{document}

\begin{frontmatter}

\title{Deep Collaborative Embedding for Information Cascade Prediction}

\author[1]{Yuhui Zhao}
\ead{zhaoyuhui@stu.scu.edu.cn}

\author[1]{Ning Yang\corref{cor1}}
\ead{yangning@scu.edu.cn}

\author[1]{Tao Lin\corref{cor1}}
\ead{lintao@scu.edu.cn}

\author[2]{Philip S. Yu}
\ead{psyu@uic.edu}

\cortext[cor1]{Ning Yang and Tao Lin share the corresponding authorship.}

\address[1]{College of Computer Science, Sichuan University, China}
\address[2]{Department of Computer Science, University of Illinois at Chicago, USA}


\begin{abstract}
Recently, information cascade prediction has attracted increasing interest from researchers, but it is far from being well solved partly due to the three defects of the existing works. First, the existing works often assume an underlying information diffusion model, which is impractical in real world due to the complexity of information diffusion. Second, the existing works often ignore the prediction of the infection order, which also plays an important role in social network analysis. At last, the existing works often depend on the requirement of underlying diffusion networks which are likely unobservable in practice. In this paper, we aim at the prediction of both node infection and infection order without requirement of the knowledge about the underlying diffusion mechanism and the diffusion network, where the challenges are two-fold. The first is what cascading characteristics of nodes should be captured and how to capture them, and the second is that how to model the non-linear features of nodes in information cascades. To address these challenges, we propose a novel model called Deep Collaborative Embedding (DCE) for information cascade prediction, which can capture not only the node structural property but also two kinds of node cascading characteristics. We propose an auto-encoder based collaborative embedding framework to learn the node embeddings with cascade collaboration and node collaboration, in which way the non-linearity of information cascades can be effectively captured. The results of extensive experiments conducted on real-world datasets verify the effectiveness of our approach.
\end{abstract}


\end{frontmatter}

%\linenumbers

\section{Introduction}
In recent years, as more and more people enjoy the services provided by Facebook, Twitter, and Weibo, etc., information cascades have become ubiquitous in online social networks, which has motivated a huge amount of researches \cite{Cheng2014Can,Li2017DeepCas,Sun2017Collaborative,Yuchen2018Influence,Siyuan2018Heterogeneous}. An important research topic is information cascade prediction, whose purpose is to predict who will be infected by a piece of information in the future \cite{Saito2008Prediction,Guille2012A,Wang2017Topological,Gao2017A}, where infection refers to the actions that users reshare (retweet) or comment a tweet, a photo, or other piece of information \cite{Bourigault2014Learning}.

While lots of methods have been proposed for information cascade prediction \cite{Saito2008Prediction,Gomez2011Inferring,Bourigault2016Representation,Xi2019IAD,Devesh2017Predicting}, the existing works often suffer from three defects. First, the existing works often focus on predicting the probability that whether a node will be infected in the future given nodes infected in the past, but ignore the prediction of infection order, i.e., which nodes will be infected earlier or later than others. However, predicting the infection order is important in many scenarios. For example, it is helpful for blocking rumor spread to know who will be the next infected node \cite{Adrien2013Information,Lan2020Containment}. Second, the existing methods often assume that information diffusion follows a parametric model such as Independent Cascade (IC) model \cite{Goldenberg2001Talk} and Susceptible-Infected (SI) model \cite{P1978The}. In real world, however, information diffusion processes are so complicated that we seldom exactly know the underlying mechanisms of how information diffuses \cite{Steeg2013Information}. At last, the existing works often assume that the explicit paths along which information propagates between nodes are observable. Yet in many scenarios we can only observe that nodes get infected but can not know who infects them \cite{Bourigault2016Representation}. For example, in viral marketing, one can track whether a customer buys a product but it is difficult to exactly determine who influences her/him.

In this paper, we aim at the problem of information cascade prediction without requirement of the knowledge about the underlying diffusion mechanism and the diffusion network. This is not easy due to the following two major challenges:

\begin{itemize}

\item \textbf{Cascading Characteristics} The probability that a node is infected by a cascade and the relative infection order mainly depend on its cascading characteristics that reveal its relation to other nodes in that cascade. The existing methods often just take into consideration the static structural properties of nodes, for example, the node neighborship in a static social network. However, the cascading characteristics of a node intuitively vary in different cascades, and different cascades can contain totally different infection ranges or orders of nodes. For example, in some cascades, one node may often get infected by certain nodes, but in other cascades, it may be more susceptible to different nodes, even though the node structural properties remain the same. Intuitively, different contents often lead to different cascading characteristics of a node and result in different underlying mechanisms in different cascades. However, in many situations it is not easy to recognize the content (i.e., what is diffused) and its underlying diffusion mechanism (i.e., why and how it is diffused). For example, we often do not know what virus is being propagated in a plague, but when and which nodes are infected can be observed. To make prediction for cascades in such situations, we have to explicitly model the observable cascading characteristics which arguably implicitly captures the effect of the unobservable content and underlying mechanism as well. Therefore, what cascading characteristics of nodes should be captured and how to capture them are crucial to our purpose.

\item \textbf{Cascading Non-linearity} Information cascades are often non-linear. The non-linearity comes from two perspectives. One is the non-linearity of the dynamics of the information cascades, and the other is the non-linearity of the structure of the social networks on which cascades exist. The non-linearity will cause the problem when nodes spread the content of a cascade, they exhibit non-linear cascading patterns (e.g., emergence pattern) that the existing shallow models can not effectively recognize. How to capture the non-linear features of nodes in information cascades is also a critical challenge for our problem.

\end{itemize}

Inspired by the impressive network representation learning ability of deep learning that has been demonstrated by the recent works \cite{Wang2016Structural,Liao2017Attributed,Chang2015Heterogeneous}, we propose a novel model called Deep Collaborative Embedding (DCE) for prediction of infection and infection order in cascades, which can learn the embeddings without assumption about the underlying diffusion model and diffusion networks. The main idea of DCE is to collaboratively embed the nodes with a deep architecture into a latent space where the closer the embeddings of the two nodes are, the more likely the two nodes will be infected in the same cascade and the closer their infection time will be.

Different from the traditional network embedding methods \cite{Wang2016Structural,Tang2015LINE,Yu2019TPNE,Perozzi2014DeepWalk}, which mainly focus on preserving the static structural properties of nodes in a network, DCE can capture not only the node structural property but also two kinds of node \textit{cascading characteristics} that are important for the prediction of node infection and infection order. One is the \textit{cascading context}, which reveals the temporal relation of nodes in a cascade. The cascading context of one node consists of two aspects, including the potential influence it receives from earlier infected nodes and their temporal relative positions in a cascade. The other kind of cascading characteristic captured by DCE is the \textit{cascading affinity}, which reveals the co-occurrence relation of nodes in cascades. Cascading affinity essentially reflects the probability that two nodes will be infected by the same cascade. Higher cascading affinity between two nodes indicates that it is more likely for them to co-occur in a cascade. Intuitively, the cascading characteristics of nodes reflect the effect of the unobservable underlying diffusion mechanisms and diffusion networks. Therefore, by explicitly preserving the node cascading characteristics, the learned embeddings also implicitly capture the effect of unobservable underlying diffusion mechanisms and diffusion network, which makes it feasible to make cascade predictions in terms of the similarity between embeddings in the latent space. As we will see later in the experiments, due to the ability to capture the cascading characteristics, the embeddings learned by DCE show a better performance in the task of infection prediction.

To effectively capture the non-linearity of information cascades, we introduce an \textit{auto-encoder based collaborative embedding} architecture for DCE. DCE consists of multi-layer non-linear transformations by which the non-linear cascading patterns of nodes can be effectively encoded into the embeddings. DCE can learn embeddings for nodes in a collaborative way, where there are two kinds of collaborations, i.e., \textit{cascade collaboration} and \textit{node collaboration}. At first, in light of the observation that a node often participates in more than one cascade of different contents, for a node DCE can collaboratively encode its cascading context features in each cascade into its embedding. In other words, the embedding of a node is learned with the collaboration of the cascades the node participates, which we call the cascade collaboration. At the same time, DCE can concurrently embed the nodes, during which the embedding for a node is generated under the constraints of its relation to other nodes, i.e., its cascading affinity to other nodes and its neighborship in social networks. In other words, the embeddings of nodes are learned with the collaboration of each other, which we call the node collaboration.

The major contributions of this paper can be summarized as follows:
\begin{enumerate}
    \item We propose a novel model called Deep Collaborative Embedding (DCE) for information cascade prediction without requirement of the knowledge about the underlying diffusion mechanism and the diffusion network. The node embeddings learned by DCE are beneficial to not only the infection prediction but also the prediction of infection order of nodes in a cascade.
    \item We propose an auto-encoder based collaborative embedding framework for DCE, which can collaboratively learn the node embeddings, preserving the node cascading characteristics including cascading context and cascading affinity, as well as the structural property.
    \item The extensive experiments conducted on real datasets verify the effectiveness of our proposed model.
\end{enumerate}

The rest of this paper is organized as follows. We give the preliminaries in Section 2. The cascading context is defined and modeled in Section 3. In section 4 we illustrate our proposed model and in Section 5 we analyze the experiments results. Finally, we briefly review the related work in Section 6 and conclude in section 7.

\section{Preliminaries and Problem Definition}

\begin{table}[t]
    \centering
    \caption{Notations}\label{glossary}
    \begin{tabular}{ll}
    \hline
    Symbol & Description\\
    \hline
    $N$ & the number of nodes\\
    $M$ & the number of cascades\\
    $\mathcal{G}$ & network\\
    $\mathcal{V}$ & the set of nodes\\
    $\mathcal{E}$ & the set of edges\\
    $\mathcal{C}$ & the set of cascades\\
%    $\mathcal{D}$ & the set of diffusion matrices\\
    $\boldsymbol{X}^{(m)}$ & the cascading context matrix of cascade $C_m$, $\boldsymbol{D}^{(m)} \in \mathbb{R}^{N\times N}$\\
    $\boldsymbol{A}$ & the cascading affinity matrix, $\boldsymbol{A} \in \mathbb{R}^{N\times N}$\\
    $\boldsymbol{S}$ & the structural proximity matrix, $\boldsymbol{S} \in \mathbb{R}^{N\times N}$\\
    $t^{(m)}_v$ & the infections time of node $v_i$ in cascade $C_m$\\
    $\boldsymbol{x}^{(m)}_v$ & the row vector of node $v$ in $\boldsymbol{X}^{(m)}$\\
    $\boldsymbol{z}_v$ & the learned embedding vector of node $v$\\
    \hline
    \end{tabular}
    \label{All Glossaries}
\end{table}

 \subsection{Basic Definitions}
We denote a social network as $\mathcal{G} = (\mathcal{V}, \mathcal{E})$, where $\mathcal{V}$ is the nodes set comprising $N$ nodes and $\mathcal{E} \subseteq \mathcal{V}\times \mathcal{V}$ is the edges set. Let $\mathcal{C} = \{ C_1, C_2, \dots, C_M\}$ be the set of $M$ information cascades. An information cascade $C_m$ ($1 \le m \le M$) observed on a social network $\mathcal{G}$ is defined as a set of timestamped infections, i.e., $C_m = \left\{ (v,  t^{(m)}_v)|v \in \mathcal{V} \wedge t^{(m)}_v < \infty \right\}$, where $(v,  t^{(m)}_v)$ represents node $v$ is infected by cascade $C_m$ at time $t^{(m)}_v$. We also say $v_i \in C_m$ if node $v_i$ participates in cascade $C_m$. Additionally, we use $C_m(t) = \{(v,  t^{(m)}_v)|v \in \mathcal{V} \wedge t^{(m)}_v < t\}$ to denote the set of nodes infected by cascade $C_m$ before time $t$, and $\overline{C}_m(t) = \mathcal{V} \backslash C_m(t)$ the set of nodes which haven't been infected before $t$. Note that the nodes in $\overline{C}_m(t)$ might or might not be infected by $C_m$ after $t$.

\subsection{Problem Definition}
The target problem of this paper can be formulated as: given a set of information cascades $\mathcal{C} = (C_1,C_2,...C_M)$ observed on a given social network $\mathcal{G}= (\mathcal{V}, \mathcal{E})$, we want to learn embeddings for nodes in $\mathcal{V}$, where the learned embeddings can preserve the cascading characteristics and structural property of nodes, so that closer embeddings indicate that the corresponding nodes are more likely to be infected by the same cascade with the closer infection time.

\section{Modeling Cascading Characteristics}
Cascading characteristics of a node reveal its relation to other nodes in information cascades, which are crucial to the prediction of node infection and infection order. In this section, we will define two kinds of cascading characteristics, the cascading context and the cascading affinity, which will be encoded into the learning embeddings.

\subsection{Cascading Context}

As mentioned before, the cascading context of a node in a cascade is supposed to capture its temporal relation to other nodes in that cascade, which includes the potential influence imposed by other nodes and their temporal infection order. There are three factors we have to consider for the definition of cascading context. First, the infection of a node is intuitively caused by the potential influence of all the nodes infected before it, and the influence declines over time. Second, the cascading context should be specific to a cascade, as one node might have different cascading contexts in different cascades. Finally, in the same cascade, the infection of one node can be influenced neither by the nodes that are infected after it, nor by the nodes that are not infected at all. Based on these ideas, we can define the cascading context as follow:
\newtheorem{definition}{Definition}
\begin{definition}
\textbf{(Cascading Context)}:
    Given the set of $M$ cascades on a social network $\mathcal{G}$ of $N$ nodes, $\mathcal{C} = (C_1,$ $C_2,...C_M)$, the cascading context of the nodes involved in cascade $C_m$ ($1 \le m \le M$) is defined as a matrix $\boldsymbol{X}^{(m)} \in \mathbb{R}^{N\times N}$. The entry at the $u$-th row and the $v$-th column of $\boldsymbol{X}^{(m)}$ represents the potential influence from node $v$ to $u$, which is defined as
     \begin{equation}
        x^{(m)}_{u,v} = \left\{
        \begin{array}{ccc}
		\exp({-\frac{t^{(m)}_u-t^{(m)}_v}{\tau}}) &,& t^{(m)}_v < t^{(m)}_u , \\
		\quad \\
		0 &,& t^{(m)}_v \geq t^{(m)}_u ,
		\end{array} \right.
     \label{Def_Temporal_Influence}
     \end{equation}
    where $t^{(m)}_u$ is the infection time of $u$ in cascade $C_m$ and $\tau$ is the decaying factor. The cascading context of node $u$ in cascade $C_m$ is defined as the row vector $\boldsymbol{x}^{(m)}_u = \boldsymbol{X}^{(m)}_{u,*}$.
\end{definition}

As we will see later, $\boldsymbol{x}^{(m)}_{u}$ will be fed into our model as it quantitatively captures $u$'s temporal relation (including the influence and the relative infection position) to the other nodes in a cascade $C_m$.

\subsection{Cascading Affinity}
As mentioned before, cascading affinity of two nodes measures the similarity of them with respect to the cascades, which can be defined in terms of their co-occurrences in historical cascades as follow:

\begin{definition}
\textbf{(Cascading Affinity)}:
    Given the set of $M$ cascades on a social network $\mathcal{G}$ of $N$ nodes,i.e., $\mathcal{C} = (C_1,$ $C_2,...C_M)$, the cascading affinity of two nodes $u$ and $v$ is represented by the entry at the $u$-th row and the $v$-th column of the cascading affinity matrix $\boldsymbol{A} \in \mathbb{R}^{N\times N}$, which is defined as the ratio of the cascades involving both $u$ and $v$, i.e.,
   \begin{equation}
    a_{u,v} = \frac{|\{C_k|u\in C_k, v\in C_k, C_k\in \mathcal{C}\}|}{|\mathcal{C}|} .
    \label{Eq_Cascade_Affinity}
    \end{equation}
    \label{Def_Cascade_Affinity}
\end{definition}

Definition \ref{Def_Cascade_Affinity} tells us that for two given nodes, the more number of cascades involving both of them, the higher their cascading affinity, and intuitively the more similar their preferences to the contents of cascades. In this sense, cascading affinity of two nodes implies that how close their embeddings should be in the latent space.

\section{Deep Collaborative Embedding}

\begin{figure}[t]
    \centering
    \epsfig{file=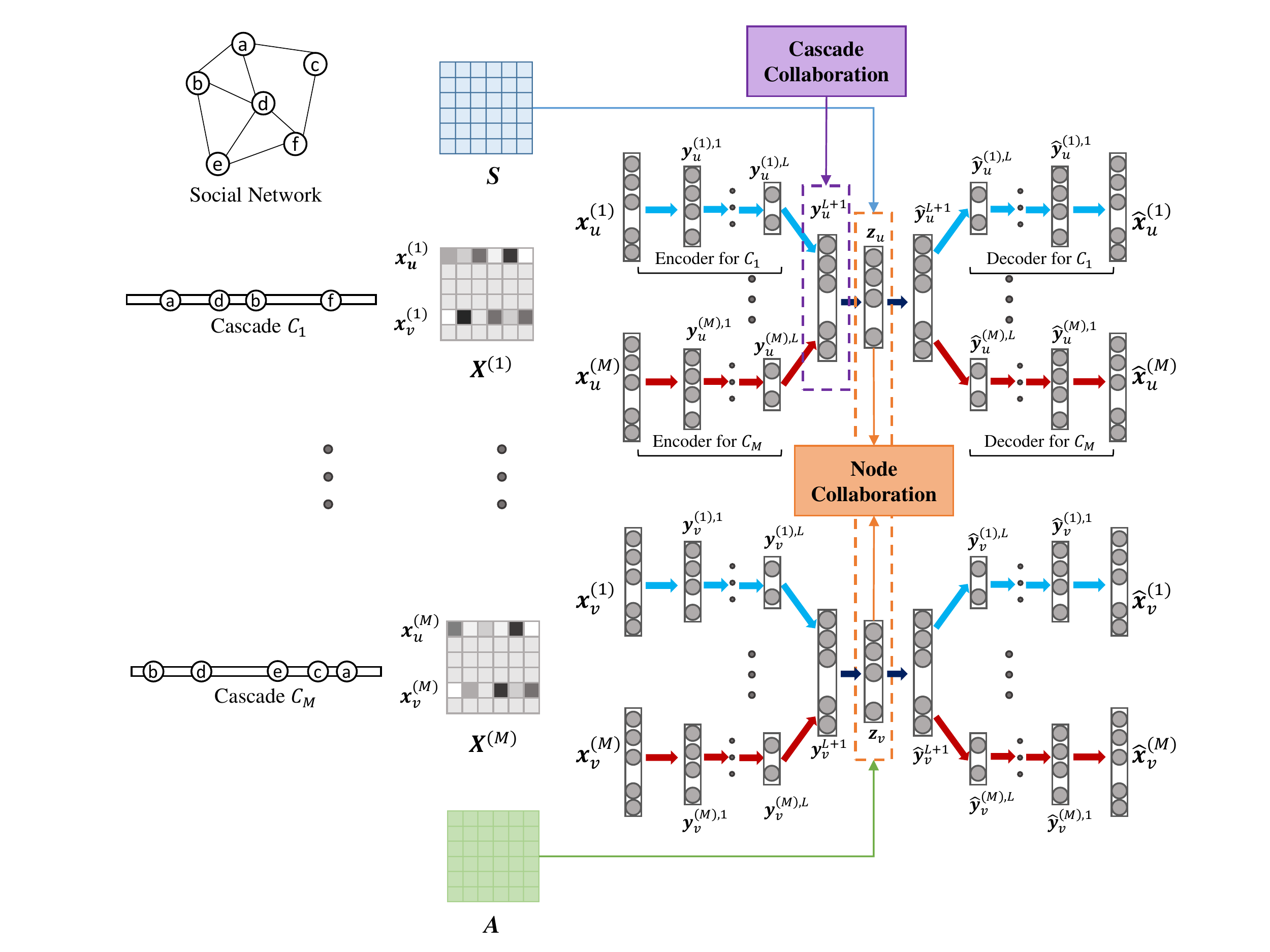, width=1\textwidth}
    \caption{Architecture of DCE}
    \label{Fig_DCE}
\end{figure}

In this paper, we propose an auto-encoder based Deep Collaborative Embedding (DCE) model, which can learn embeddings for nodes in a given social network, based on the $M$  cascades $C_1, \dots, C_M$ observed on the network, so that the learned embeddings can be used for cascade prediction without knowing the underlying diffusion mechanisms and the explicit diffusion networks. In this section, we first present the architecture of the Deep Collaborative Embedding (DCE) model in detail, and then we describe the objective function and the learning of DCE.

\subsection{Architecture of DCE}
The architecture of DCE is shown in Fig.\ref{Fig_DCE}. As we can see from Fig.\ref{Fig_DCE}, DCE learns the embeddings through two collaborations, the cascade collaboration and the node collaboration. With the cascade collaboration, DCE can generate the result $d$-dimensional embedding $\boldsymbol{z}_v \in \mathbb{R}^d$ for a node $v$ by collaboratively encoding its $M$ cascading contexts, $\boldsymbol{x}^{(m)}_v$ ($1 \le m \le M$). At first, DCE will learn $M$ intermediate embeddings $\boldsymbol{y}^{(1)}_v, \dots, \boldsymbol{y}^{(M)}_v$ for $v$ by $M$ auto-encoders, respectively, each of which corresponds to a cascade. The auto-encoder for cascade $C_m$ ($1 \le m \le M$) takes the $v$'s cascading context $\boldsymbol{x}^{(m)}_v$ in the cascade $C_m$ as input, and then generates the intermediate embedding of $v$ in cascade $C_m$, $\boldsymbol{y}^{(m)}_v$, through its encoder part consisting of $L$ non-linear hidden layers defined by the following equations:
\begin{equation}
\begin{aligned}
\boldsymbol{y}^{(m),1}_{v} &=\sigma\left(\boldsymbol{W}^{(m),1} \boldsymbol{x}^{(m)}_{v}+ \boldsymbol{b}^{(m),1}\right),\\
\boldsymbol{y}^{(m),l}_{v} &=\sigma\left(\boldsymbol{W}^{(m),l} \boldsymbol{y}^{(m),l-1}_{v} + \boldsymbol{b}^{(m),l}\right),\quad \forall l \in \{2,3,...L \},
\end{aligned}
\label{Eq_Encoder}
\end{equation}
where $\boldsymbol{y}^{(m),l}_{v}$ is the output vector of $l$-th hidden layer of $m$-th auto-encoder taking $\boldsymbol{x}^{(m)}_{v}$ as input, $\boldsymbol{W}^{(m),l}$ is the parameter matrix of that layer, and $\boldsymbol{b}^{(m),l}$ is the corresponding bias.

At last, the result embedding $\boldsymbol{z}_{v}$ is generated by fusing the $M$ intermediate embeddings $\boldsymbol{y}^{(m), L}_{v}$ ($1 \le m \le M$) through the following non-linear mappings:
\begin{equation}
\begin{aligned}
\boldsymbol{y}^{L+1}_{v} &=\sigma\bigg(\sum_{m = 1}^M(\boldsymbol{W}^{(m),L+1} \boldsymbol{y}^{(m),L}_{v}+ \boldsymbol{b}^{(m),L+1})\bigg),\\
\boldsymbol{z}_{v}\quad &=\sigma\left(\boldsymbol{W}^{L+2} \boldsymbol{y}^{L+1}_{v} + \boldsymbol{b}^{L+2}\right).
\end{aligned}
\label{Eq_Z}
\end{equation}
Symmetrically, the decoder part of the auto-encoder for cascade $C_m$ is defined by the following equations:
\begin{equation}
\begin{aligned}
\hat{\boldsymbol{y}}^{L+1}_{v} &=\sigma\left(\hat{\boldsymbol{W}}^{L+2} \boldsymbol{z}_{v} + \hat{\boldsymbol{b}}^{L+2}\right),\\
\hat{\boldsymbol{y}}^{(m),L}_v &=\sigma\left(\hat{\boldsymbol{W}}^{(m),L+1} \hat{\boldsymbol{y}}^{L+1}_{v} + \hat{\boldsymbol{b}}^{(m),L+1}\right),\\
\hat{\boldsymbol{y}}^{(m),l-1}_{v} &=\sigma\left(\hat{\boldsymbol{W}}^{(m),l-1} \hat{\boldsymbol{y}}^{(m),l}_{v} + \hat{\boldsymbol{b}}^{(m),l-1}\right),\quad \forall l \in \{2,3,...L \},\\
\hat{\boldsymbol{x}}^{(m)}_{v} &=\sigma\left(\hat{\boldsymbol{W}}^{(m),1} \hat{\boldsymbol{y}}^{(m),1}_{v}+ \hat{\boldsymbol{b}}^{(m),1}\right).
\end{aligned}
\label{Eq_Decoder}
\end{equation}
In the above Equations (\ref{Eq_Encoder}), (\ref{Eq_Z}), and (\ref{Eq_Decoder}), the parameter matrices $\boldsymbol{W}$ and $\hat{\boldsymbol{W}}$, and the bias vectors $\boldsymbol{b}$ and $\hat{\boldsymbol{b}}$ are the parameters that will be learned from training data.

At the same time, with the node collaboration, DCE can concurrently embed the nodes into latent space, by which the similarity between nodes in the social network can be captured into the learned embeddings. Particularly, to regulate the closeness between any two embeddings $\boldsymbol{z}_{u}$ and $\boldsymbol{z}_v$, DCE will impose the constraints of the cascading affinity $a_{u,v}$ and structural proximity $s_{u,v}$ between $u$ and $v$ via Laplacian Eigenmaps, which will be described in detail in next subsection.

\subsection{Optimization Objective of DCE}
\subsubsection{Loss Function for Cascade Collaboration}
At first, as described in last subsection, $M$ auto-encoders defined by Equations (\ref{Eq_Encoder}), (\ref{Eq_Z}), and (\ref{Eq_Decoder}) fulfill the cascade collaboration for embedding $v$ by reconstructing its $M$ cascading contexts $\boldsymbol{x}_v^{(m)}$. The optimization objective for this part is to minimize the reconstruction error between $\boldsymbol{x}_v^{(m)}$ and $\hat{\boldsymbol{x}}_i^{(m)}$, of which the loss function is defined as follow:
\begin{equation}
\begin{aligned}
\mathcal{L}_x &= \sum_{m=1}^{M}\sum_{v\in \mathcal{V}}\big\|(\boldsymbol{x}^{(m)}_{v}-\hat{\boldsymbol{x}}^{(m)}_{v})\big\|^2_2 \quad\\
&= \sum_{m}^{M}\big\|(\boldsymbol{X}^{(m)}-\hat{\boldsymbol{X}}^{(m)})\big\|^2_F,
\end{aligned}
\label{Eq_Lx}
\end{equation}
where $\boldsymbol{X}^{(m)}$ and $\hat{\boldsymbol{X}}^{(m)}$ are the original cascading context matrix and the reconstructed cascading context matrix of cascade $C_m$, respectively, which are defined in Definition \ref{Def_Temporal_Influence}.

The cascading context vectors $\boldsymbol{x}^{(m)}_v$ are often sparse, which may leads to undesired $\boldsymbol{0}$ vectors in the embeddings $\boldsymbol{z}_v$ and the reconstructed $\boldsymbol{x}^{(m)}_v$ if the sparse vectors $\boldsymbol{x}^{(m)}_{v}$ are straightforwardly fed into DCE. To overcome this issue, inspired by the idea used in the existing works \cite{Wang2016Structural,Zhang2017BL} which assign more penalty (corresponding to larger weight) to the loss incurred by non-zero elements than that incurred by zero elements, the $\mathcal{L}_x$ can be redefined as
\begin{equation}
\begin{aligned}
\mathcal{L}_x &= \sum_{m}^{M}\sum_{v\in \mathcal{V}}\big\|(\boldsymbol{x}^{(m)}_{v}-\hat{\boldsymbol{x}}^{(m)}_{v}) \odot \boldsymbol{p}^{(m)}_{v}\big\|^2_2 \\
&= \sum_{m}^{M}\big\|(\boldsymbol{X}^{(m)}-\hat{\boldsymbol{X}}^{(m)}) \odot \boldsymbol{P}^{(m)}\big\|^2_F,
\end{aligned}
\label{Eq_Lx}
\end{equation}
where $\odot$ denotes the Hadamard product, and the $u$-th column vector of the matrix $\boldsymbol{P}^{(m)} \in \mathbb{R}^{N \times N}$ is the weight vector $\boldsymbol{p}^{(m)}_{u} = \{p^{(m)}_{u,v}\}_{v\in \mathcal{V}}$ assigned to cascading context $\boldsymbol{x}_u^{(m)}$. An entry $p^{(m)}_{u,v} = \rho > 1$ if $x^{(m)}_{u,v} \ne 0$, otherwise $p^{(m)}_{u,v} = 1$.

\subsubsection{Loss Functions for Node Collaboration}

Next we introduce the loss function for node collaboration. As mentioned in last subsection, through the node collaboration the embeddings $\boldsymbol{z_i}$ will preserve the cascading affinity of nodes in cascades and the structural proximity of nodes in social network. Following the idea of Laplacian Eigenmaps, we weight the similarity between two embeddings with the cascading affinity of their corresponding nodes, which leads to the following loss function:
\begin{equation}
\mathcal{L}_a = \sum_{u,v \in \mathcal{V}} a_{u,v}\big\|\boldsymbol{z}_u-\boldsymbol{z}_v\big\|^2_2,
\label{Eq_La}
\end{equation}
where $a_{u,v}$ is the cascading affinity between $u$ and $v$ defined in Equation (\ref{Eq_Cascade_Affinity}). The insight of Equation (\ref{Eq_La}) is that a penalty will be imposed when two nodes with high cascading affinity are relocated far away in the latent space.

Similarly, we also weight the similarity between two embeddings with the structural proximity of their corresponding nodes, which leads to the following loss function:
\begin{equation}
\mathcal{L}_s = \sum_{u,v\in \mathcal{V}} s_{u,v}\big\|\boldsymbol{z}_u-\boldsymbol{z}_v\big\|^2_2,
\label{Eq_Ls}
\end{equation}
where $s_{u,v}$ is the structural proximity between $u$ and $v$ in social network. Note that it does not matter how to define $s_{u,v}$, and theoretically, the node structural proximity of any order can be used for $s_{u,v}$. In this paper, we employ the first-order proximity \cite{Tang2015LINE} to define $s_{u,v}$.  To be more specific, $s_{u,v} = 1$ if $u$ and $v$ are connected by a link in the network, otherwise $s_{u,v} = 0$.

Let $\boldsymbol{L}^{(a)}$ be the laplacian matrix of the cascading affinity matrix $\boldsymbol{A}$, i.e., $\boldsymbol{L}^{(a)} = \boldsymbol{D}^{(a)} - \boldsymbol{A}$, where $\boldsymbol{D}^{(a)}$ is diagonal and $\boldsymbol{D}^{(a)}_{u,u} = \sum^{N}_{v}a_{u,v}$. Let $\boldsymbol{S}$ be the structural proximity matrix whose entry at $u$-th row and $v$-th column is $s_{u, v}$, and similarly, let $\boldsymbol{L}^{(s)}$ be its laplacian matrix, i.e., $\boldsymbol{L}^{(s)} = \boldsymbol{D}^{(s)} - \boldsymbol{A}$, where $\boldsymbol{D}^{(s)}$ is also diagonal and $\boldsymbol{D}^{(s)}_{u,u} = \sum_{v\in \mathcal{V}}s_{u,v}$. Then we can rewrite the Equations (\ref{Eq_La}) and (\ref{Eq_Ls}) with their matrix forms:
\begin{equation}
\mathcal{L}_a = 2tr(\boldsymbol{Z}^T\boldsymbol{L}^{(a)}\boldsymbol{Z}),
\label{Eq_La_Matrix}
\end{equation}
and
\begin{equation}
\mathcal{L}_s = 2tr(\boldsymbol{Z}^T\boldsymbol{L}^{(s)}\boldsymbol{Z}),
\label{Eq_Ls_Matrix}
\end{equation}
where $\boldsymbol{Z}$ is the embedding matrix whose $i$-th column is $\boldsymbol{z}_i$.

\subsubsection{The Complete Loss Function}
By combining $\mathcal{L}_x$, $\mathcal{L}_a$, and $\mathcal{L}_s$, we can define the $\textbf{complete loss function}$ of DCE as follow:
\begin{equation}
\mathcal{L} =\mathcal{L}_x + \alpha\mathcal{L}_a + \beta\mathcal{L}_s + \gamma\mathcal{L}_{reg},
\label{Eq_TL}
\end{equation}
where $\mathcal{L}_{reg} = \sum_{l}^{L+2}\sum_{m}^{M}(\big\|\boldsymbol{W}^{(m), l}\big\|^2_2 + \big\|\hat{\boldsymbol{W}}^{(m), l}\big\|^2_2$ is a $\mathcal{L}$2-norm regularizer term to avoid overfitting, and $\alpha$, $\beta$, and $\gamma$ are nonnegative parameters used to control the contributions of the terms.

%DCE Algorithm
\renewcommand{\algorithmicrequire}{\textbf{Input:}}
\renewcommand{\algorithmicensure}{\textbf{Output:}}
\begin{algorithm}[t]
  \caption{learning algorithm of DCE}
  \label{alg:A}
  \begin{algorithmic}[1]
    \REQUIRE ~~ \\
       The set of cascading context matrices $\mathcal{X} = (\boldsymbol{X}_1, \boldsymbol{X}_2,...\boldsymbol{X}_M)$, cascading affinity matrix $\boldsymbol{A}$, structural proximity matrix $\boldsymbol{S}$, and the parameters $\alpha, \beta$ and $\gamma$. \\
    \ENSURE ~~ \\
       Node embeddings $\boldsymbol{Z}$.

    \STATE Initialize parameters $\boldsymbol{W}$, $\hat{\boldsymbol{W}}$, $\boldsymbol{b}$, and $\hat{\boldsymbol{b}}$.
    \REPEAT
      \STATE Compute $\boldsymbol{Z}, \hat{\boldsymbol{X}}$ according to Equations (\ref{Eq_Encoder}), (\ref{Eq_Z}) and (\ref{Eq_Decoder}).
      \STATE Compute total loss $\mathcal{L}$ according to Equation (\ref{Eq_TL}).
      \STATE Update $\boldsymbol{W}$, $\hat{\boldsymbol{W}}$, $\boldsymbol{b}$, and $\hat{\boldsymbol{b}}$ according to Equations (\ref{Eq_W}) to (\ref{Eq_b_Hat}) using SGD.
    \UNTIL $\mathcal{L}$ converges.
  \end{algorithmic}
\end{algorithm}

\subsection{Learning of DCE}
DCE model can be learned using Stochastic Gradient Descent (SGD), the gradients of which are given by the follow equations:

\begin{flalign}
\begin{split}
\frac{\partial\mathcal{L}}{\partial\boldsymbol{W}^{(m),l}} &= \sum_{m}^{M}2(\hat{\boldsymbol{X}}^{(m)} - \boldsymbol{X}^{(m)}) \odot \boldsymbol{P}^{(m)} \cdot \frac{\partial\hat{\boldsymbol{X}}^{(m)}}{\partial\boldsymbol{W}^{(m),l}} + \gamma \cdot  \frac{\boldsymbol{W}^{(m),l}}{2} + \quad\\
&\quad\alpha \cdot (2(\boldsymbol{L}^{(a)} + {\boldsymbol{L}^{(a)}}^T)\cdot\boldsymbol{Z}) \cdot \frac{\partial\boldsymbol{Z}}{\partial\boldsymbol{W}^{(m),l}} + \\
&\quad\beta \cdot (2(\boldsymbol{L}^{(s)} + {\boldsymbol{L}^{(s)}}^T)\cdot\boldsymbol{Z}) \cdot \frac{\partial\boldsymbol{Z}}{\partial\boldsymbol{W}^{(m),l}},\\
\end{split}
\quad
\label{Eq_W}
\end{flalign}

\begin{flalign}
\begin{split}
\frac{\partial\mathcal{L}}{\partial\boldsymbol{b}^{(m),l}} &= \sum_{m}^{M}2(\hat{\boldsymbol{X}}^{(m)} - \boldsymbol{X}^{(m)}) \odot \boldsymbol{P}^{(m)} \cdot \frac{\partial\hat{\boldsymbol{X}}^{(m)}}{\partial\boldsymbol{b}^{(m),l}} + \gamma \cdot  \frac{\boldsymbol{b}^{(m),l}}{2} + \qquad\\
&\quad\alpha \cdot (2(\boldsymbol{L}^{(a)} + {\boldsymbol{L}^{(a)}}^T)\cdot\boldsymbol{Z}) \cdot \frac{\partial\boldsymbol{Z}}{\partial\boldsymbol{b}^{(m),l}} + \\
&\quad\beta \cdot (2(\boldsymbol{L}^{(s)} + {\boldsymbol{L}^{(s)}}^T)\cdot\boldsymbol{Z}) \cdot \frac{\partial\boldsymbol{Z}}{\partial\boldsymbol{b}^{(m),l}},\\
\end{split}
\quad
\label{Eq_b}
\end{flalign}

\begin{flalign}
\begin{split}
\frac{\partial\mathcal{L}}{\partial\hat{\boldsymbol{W}}^{(m),l}} &= \sum_{m}^{M}2(\hat{\boldsymbol{X}}^{(m)} - \boldsymbol{X}^{(m)}) \odot \boldsymbol{P}^{(m)} \cdot \frac{\partial\hat{\boldsymbol{X}}^{(m)}}{\partial\hat{\boldsymbol{W}}^{(m),l}} + \gamma \cdot  2\hat{\boldsymbol{W}}^{(m),l},\\
\end{split}
\quad
\label{Eq_W_Hat}
\end{flalign}

\begin{flalign}
\begin{split}
\frac{\partial\mathcal{L}}{\partial\hat{\boldsymbol{b}}^{(m),l}} &= \sum_{m}^{M}2(\hat{\boldsymbol{X}}^{(m)} - \boldsymbol{X}^{(m)}) \odot \boldsymbol{P}^{(m)} \cdot \frac{\partial\hat{\boldsymbol{X}}^{(m)}}{\partial\hat{\boldsymbol{b}}^{(m),l}} + \gamma \cdot  2\hat{\boldsymbol{b}}^{(m),l},\\
\end{split}
\qquad
\label{Eq_b_Hat}
\end{flalign}
where the partial derivatives on the right side of the equations can be computed using back-propagation.

The learning process is given in Algorithm \ref{alg:A}. Note that in each iteration, the parameters are updated (Line 5) once the embeddings $\boldsymbol{z}_v$, $v \in \mathcal{V}$ are concurrently generated (Line 3). Such concurrent embedding scheme ensures the cascading context can be encoded into the embeddings as well as the cascading affinity and the structural proximity of nodes can be preserved at the same time.

\section{Experiments}
In this section, we will present the details of experiments conducted on real-world datasets. The experiments include two parts, the tuning of the hyper-parameters and the verifying of DCE. Particularly, to verify the effectiveness of DCE, we will check whether the embeddings learned by DCE improve the performance of the prediction of information cascades on the real world datasets.

  \subsection{Settings}
    \subsubsection{Datasets}
We verify the effectiveness of our method through experiments conducted on three real datasets, Digg, Twitter, and Weibo, which are described as follows:

\textbf{Digg} is a website where users can submit stories and vote for the stories they like \cite{Hogg2012Social}. The dataset extracted from Digg contains 3,553 stories, 139,409 users, and 3,018,197 votes with timestamps. A vote for a story is treated as an infection of that story, and the votes for the same story constitute a cascade. In addition, a social link exists between two users if one of them is watching or is a fan of the other one.

\textbf{Twitter} is a social media network which offers microblog service \cite{Weng2013Virality}. The dataset extracted from Twitter comprises 510,795 users and 12,054,205 tweets with timestamps, where each tweet is associated with a hashtag. If the hashtag is adopted in one user's tweet, we consider it infects that user. The tweets sharing the same hashtag are treated as a cascade, and 1,345,913 cascades are contained in the dataset. In addition, the users are linked by their following relationships.

\textbf{Weibo} is a Twitter-like social network \cite{Zhang2013Social}. The dataset extracted from Weibo contains 1,340,816 users and their 31,444,325 tweets with timestamps. A retweeting action of a user is viewed as an infection of the retweeted tweet to that user. The retweetings of the same tweet constitute a cascade, and the dataset contains 232,978 cascades of different tweets. The users in Weibo network are also connected by following relationships.

The statistics of the datasets are summarized in Table \ref{Tab_DS}. On each dataset, we randomly select 60\% of the total cascades as training set, 20\% as validating set, and the remaining 20\% as testing set.

\begin{table}[t]
\centering
\small
\caption{The statistics of datasets}
\begin{tabular}{p{1.0cm}<{\centering}|p{1.2cm}<{\centering}p{1.3cm}<{\centering}p{1.0cm}<{\centering}p{1.3cm}<{\centering}p{1.3cm}<{\centering}p{2.1cm}<{\centering}}
\hline
Dataset & \#Nodes\quad & \#Links\quad & \ Avg. Degree\quad & \#Cascades\quad & \#Infections\quad & \ Avg. Cascade Length\quad \\
\hline
Digg & 139,409 & 1,731,658 & 12.4 & 3,553 & 3,018,197 & 849.5\\
Twitter & 510,795 & 14,273,311 & 27.9 & 1,345,913 & 12,054,205 & 9.0\\
Weibo & 1,340,816 & 308,489,739 & 230.1 & 232,978 & 31,444,325 & 135.0\\
\hline
\end{tabular}
\label{Tab_DS}
\end{table}

\subsubsection{Baselines}
In order to demonstrate the effectiveness of DCE, we compare it with the following baseline methods:

\textbf{NetRate} NetRate is a generative cascade model which exploits infection times of nodes without assumptions on the network structure \cite{Rodriguez2011Uncovering}. It models information diffusion process as discrete networks of continuous temporal process occurring at different rates, and then infers the edges of the global diffusion network and estimates the transmission rates of each edge that best explain the observed data.

\textbf{CDK} CDK maps nodes participating in information cascades to a latent representation space using a heat diffusion process \cite{Bourigault2014Learning}. It treats learning diffusion as a ranking problem and learns heat diffusion kernels that defines, for each node of the network, its likelihood to be reached by the diffusing content, given the initial source of diffusion. Here we adopt the without-content version of CDK considering that other baselines and our approach are not designed to deal with diffusion content.

\textbf{Topo-LSTM} Topo-LSTM uses directed acyclic graph as the diffusion topology to explore the diffusion structure of cascades rather than regarding it as merely a sequence of nodes ordered by their infection timestamps \cite{Wang2017Topological}. Then it puts dynamic DAGs into a LSTM-based model to generate topology-aware embeddings for nodes as outputs.  The infection probability at each time step will be computed according to the embeddings.

\textbf{Embedded-IC} Embedded-IC is a representation learning technique for inference of Independent Cascade (IC) model \cite{Bourigault2016Representation}. Embedded-IC can embed users in cascades into a latent space and infer the diffusion probability between users based on the relative positions of the users in the latent space.

\textbf{DCE-C} DCE-C is a special version of the proposed DCE, where the node collaborations of cascading affinity and structural proximity are removed while only the cascade collaboration of cascading contexts is kept.

\subsection{Cascade Prediction}
In this paper, we evaluate the learned embeddings by applying them to the task of information cascade prediction, the details of which are described as follows. %As in DCE the embeddings are learned collaboratively based on different cascades, we merge all testing cascades as one testing cascade $C$ to perform prediction task, with ignoring which cascade each node belongs to and ordering the nodes by their infection time.

For a testing cascade $C$, given a set of seed nodes which are infected before, we predict the infection probabilities for the remaining nodes and their infecting order. To be more specific, the size of the seed set will be $1\%$ of the total number of the nodes. Let $V_t \subset \mathcal{V}$ be the set of nodes that are predicted before time step $t+1$, and then the probability that one node $u \in \mathcal{V} \backslash V_t$ will be infected at $t+1$ is
\begin{equation}
P(u|V_t) = 1 - \prod_{v \in V_t}\big(1-P(u|v)\big),
\label{Eq_Prob_Infection}
\end{equation}
where $P(u|v)$ is the probability that $u$ is infected by $v$. Our idea of computing $P(u|v)$ is based on the similarity between the embeddings, which is defined as
\begin{equation}
P(u|v) = \frac{1}{1 + \exp{(\big\|\boldsymbol{z}_{v}-\boldsymbol{z}_{u}\big\|^2_2)}},
\label{Eq_PRO}
\end{equation}
where $\boldsymbol{z}_{u}$ and $\boldsymbol{z}_{v}$ are embedding vectors of nodes $u$ and $v$, respectively, and the similarity is measured by Euclidean distance. For each uninfected node $u \in \overline{C}(t)$, its infection probability can be computed according to Equation (\ref{Eq_Prob_Infection}), and we can obtain a list $\hat{R}_C$ of the nodes in descending order of their infection probabilities. Comparing $\hat{R}_C$ with the ground truth $R_C$, we can evaluate the performance of the prediction with two metrics, Mean Average Precision(MAP) and order-Precision.

As a metric originating from information retrieval, MAP can evaluate the prediction of information cascades by taking positions of nodes in the predicting list into consideration. We first define the top-$n$ precision of $\hat{R}_C$ as the hit rate of the first $n$ nodes of $\hat{R}_C$ over the ground truth, i.e.,
\begin{equation}
p_{C, n} = \frac{|\hat{R}_{C,n} \cap R_C|}{|\hat{R}_{C,n}|},
\label{Eq_q}
\end{equation}
where $\hat{R}_{C,n}$ is the set of first $n$ nodes of $\hat{R}_C$. Then based on $p_{C, n}$, we can define the average precision of $\hat{R}_C$ as
\begin{equation}
AP_{C} =  \frac{\sum_{v \in R_C} {p_{C,r_{c, v}}}}{|R_C|},
\label{Eq_Q}
\end{equation}
where $r_{c, v}$ denotes the rank of node $v$ in $\hat{R}_C$ and $p_{C,r_{c, v}}$ is the top-$r_{c, v}$ precision of $\hat{R}_C$. From Equations (\ref{Eq_q}) and (\ref{Eq_Q}) we can see that, it will lead to a low $AP_{C}$ if too many nodes which occur in $R_C$ but rank low in $\hat{R}_C$. What's more, we set the size $k$ of the predicted list $\hat{R}_C$ in \{100, 300, 500, 700, 900\} to compute $AP_C@k$ among the first $k$ nodes. Finally, $MAP@k$ can be defined as the average of $AP_C@k$ over testing set $\mathcal{C}_t$, i.e.,
\begin{equation}
MAP@k = \frac{1}{|\mathcal{C}_t|}\sum_{C \in \mathcal{C}_t}AP_C@k.
\end{equation}
%What's more, we set the size of the predicted list $\hat{R}_C$ in \{100, 300, 500, 700, 900\} to compute MAP@$k$ among the first $k$ nodes.

To evaluate the prediction of infection order, we propose a new metric, order-Precision, which is defined as
\begin{equation}
P_o = \frac{1}{|\mathcal{C}_t|}\sum_{C \in \mathcal{C}_t}\frac{1}{|R_C|}\sum_{v \in R_C \cap \hat{R}_C}\frac{|\hat{R}_C(\hat{t}_v^C) \cap R_C(t_v^{C})|}{|\hat{R}_C(\hat{t}_v^C) \cap R_C|},
\label{Eq_Order_Precision}
\end{equation}
where $t_v^C$ is the true infection time of $v$ and $\hat{t}_v^{C}$ is the predicted one, and $R_C(t_v^C)$ and $\hat{R}_C(\hat{t}_v^{C})$ denotes the sets of nodes infected before node $v_v$ in the ground truth list and the predicted list respectively. The idea of Equation (\ref{Eq_Order_Precision}) is that the more nodes with more similar relative orders of nodes in $R_C$ and $\hat{R}_C$, the higher the order-Precision of $\hat{R}_C$. First, to evaluate the similarity of node $v'$s relative orders in $R_C$ and $\hat{R}_C$, we consider a heuristic indicator, the number of the nodes that are infected before node $v$ and shared by $R_C$ and $\hat{R}_C$, i.e., $|\hat{R}_C(\hat{t}_v^C) \cap R_C(t_v^{C})|$, and the larger this number is, the more similar the relative orders will be. Then we can obtain the relative order similarity for one single testing cascade $C$ by taking the average over all nodes shared by $R_C$ and $\hat{R}_C$. Finally, the overall order-Precision is the average of the relative order similarities over all testing cascades in $\mathcal{C}_t$.

\begin{figure}
    \begin{minipage}[t]{0.5\textwidth}
        \centering
         \subfigure[MAP]{\includegraphics[width=1\textwidth]{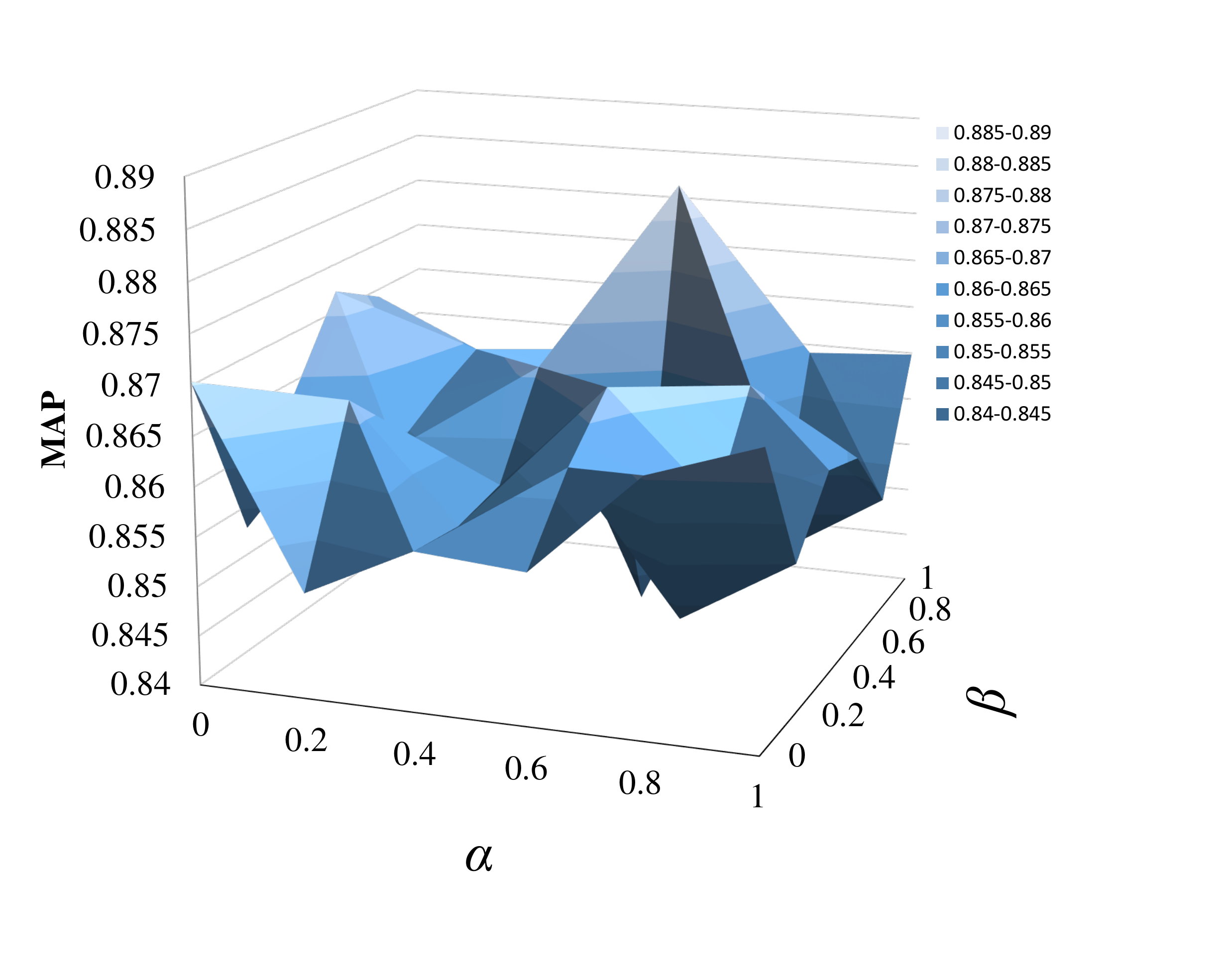}}
    \end{minipage}
    \begin{minipage}[t]{0.5\textwidth}
        \centering
         \subfigure[order-Precision]{\includegraphics[width=1\textwidth]{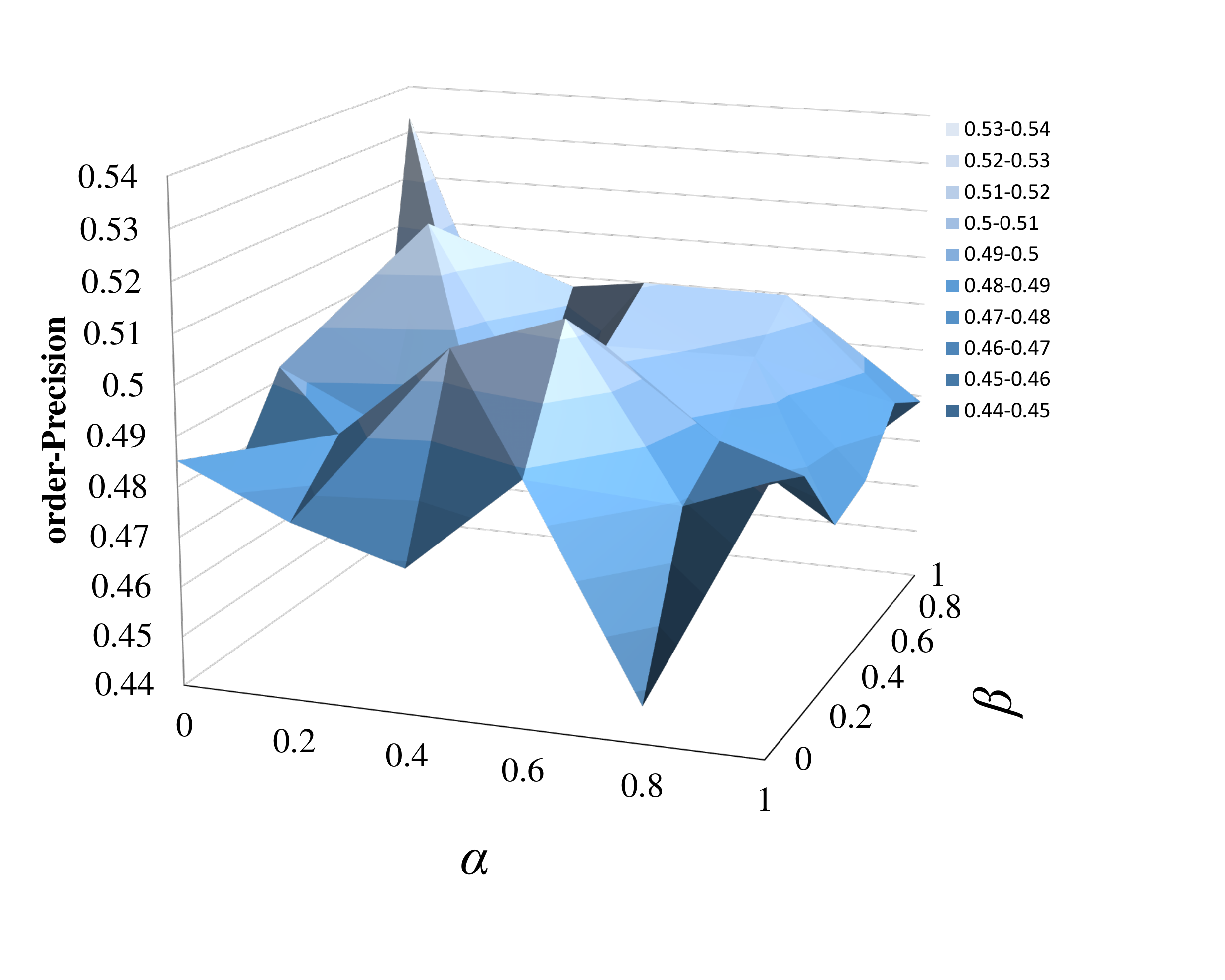}}
    \end{minipage}
    \caption{Tuning the parameter $\alpha$ and $\beta$ on Digg.}
    \label{ParaInDigg}

    \begin{minipage}[t]{0.5\textwidth}
        \centering
         \subfigure[MAP]{\includegraphics[width=1\textwidth]{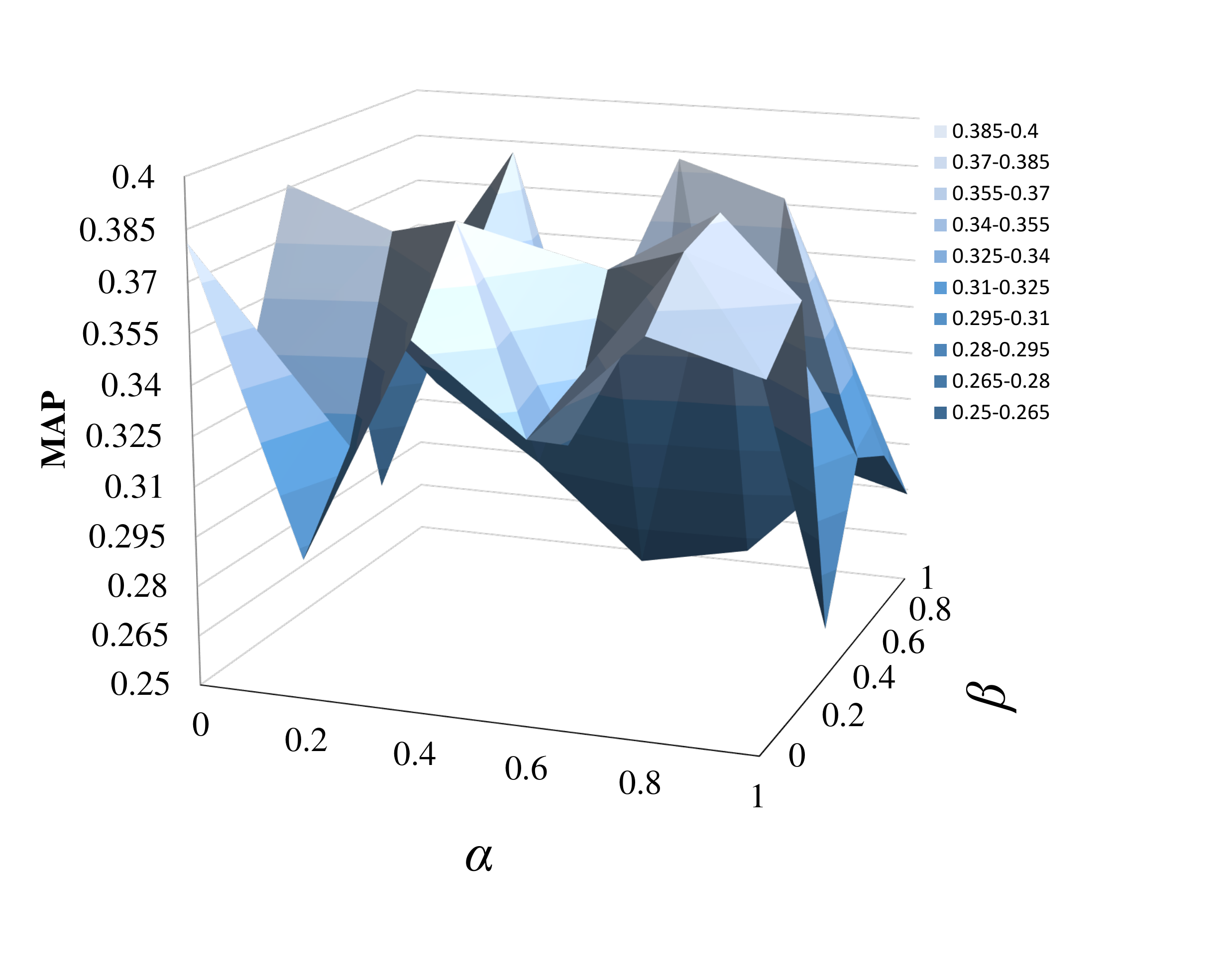}}
    \end{minipage}
    \begin{minipage}[t]{0.5\textwidth}
        \centering
         \subfigure[order-Precision]{\includegraphics[width=1\textwidth]{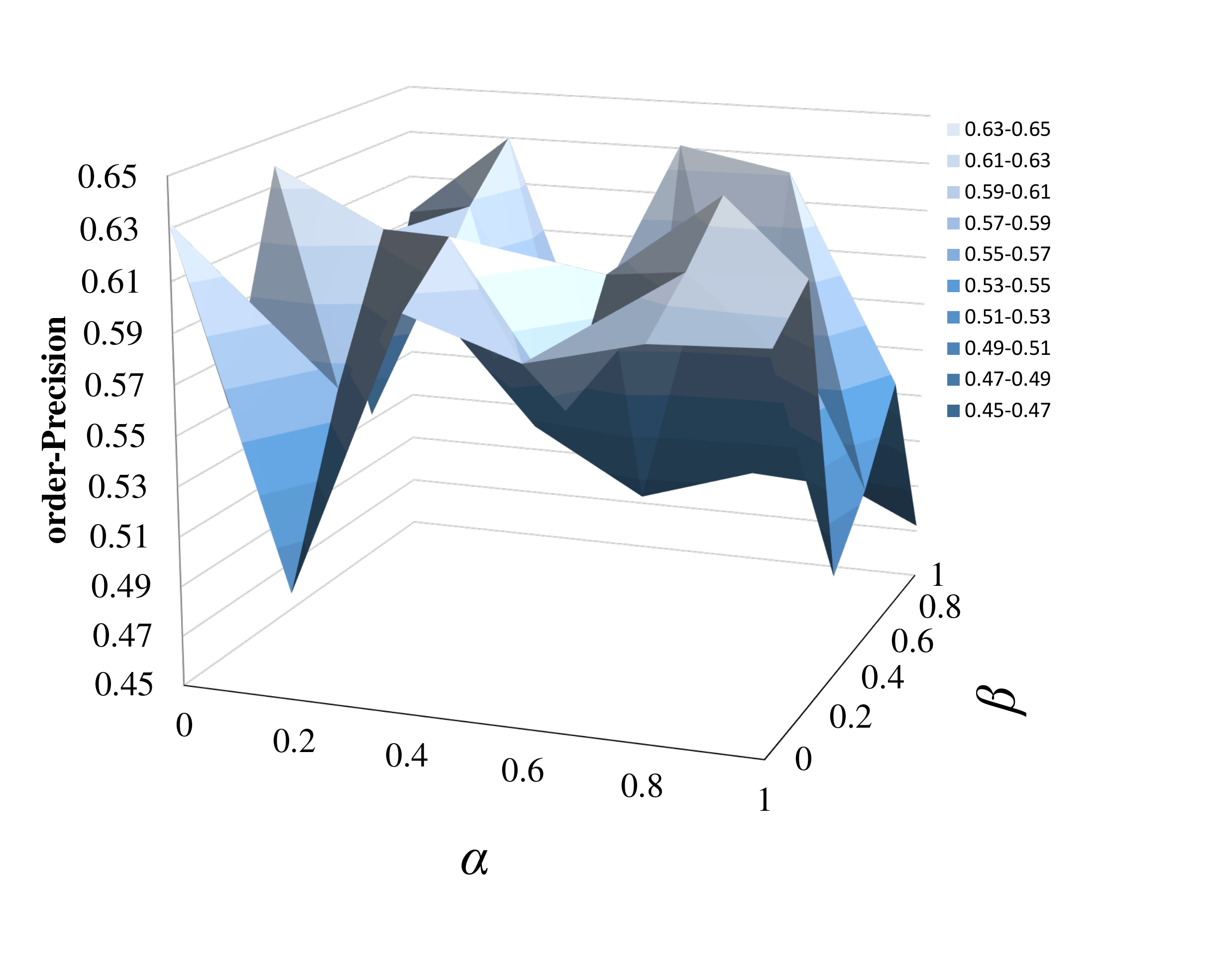}}
    \end{minipage}
    \caption{Tuning the parameter $\alpha$ and $\beta$ on Twitter.}
    \label{ParaInTwitter}

    \begin{minipage}[t]{0.5\textwidth}
        \centering
         \subfigure[MAP]{\includegraphics[width=1\textwidth]{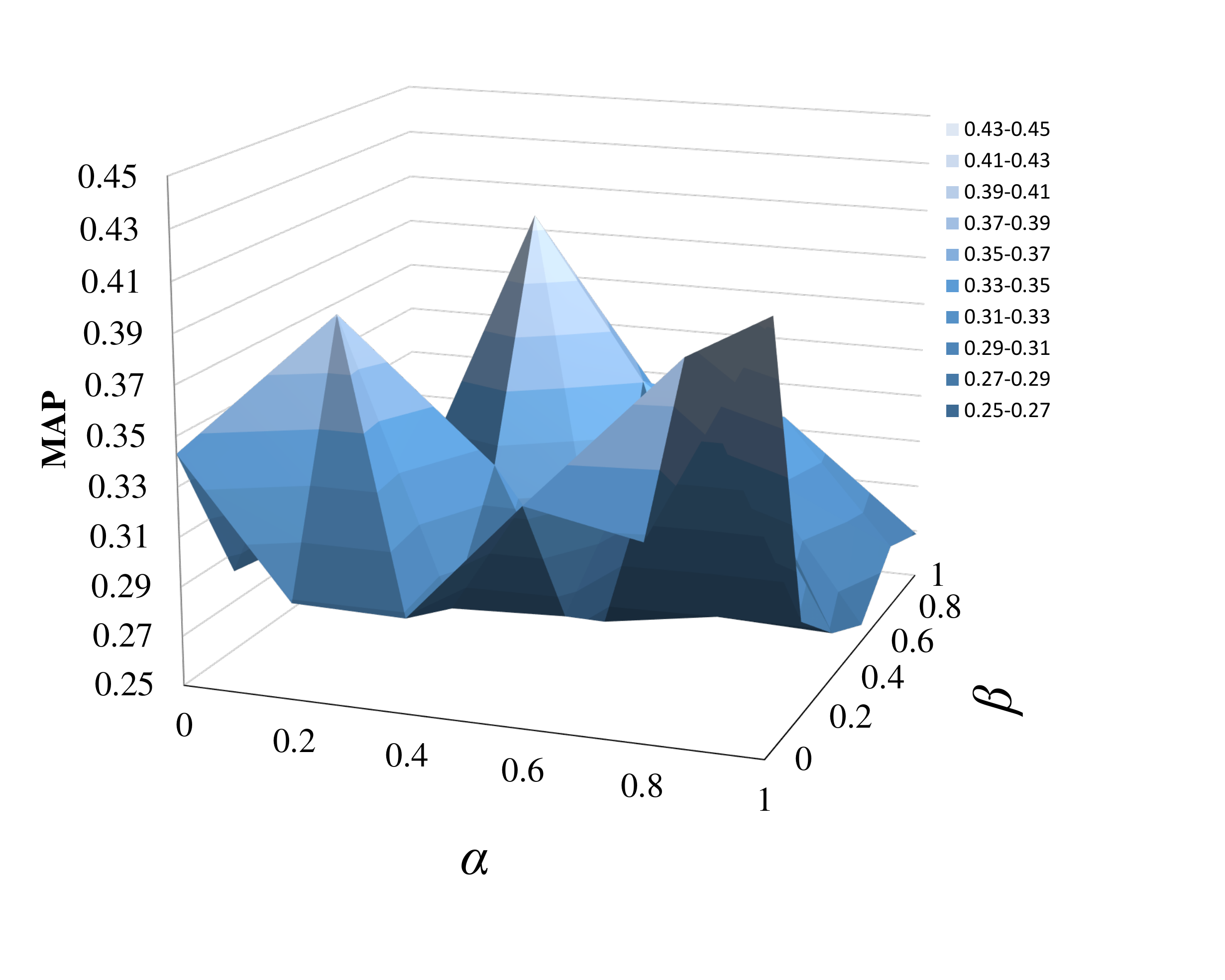}}
    \end{minipage}
    \begin{minipage}[t]{0.5\textwidth}
        \centering
         \subfigure[order-Precision]{\includegraphics[width=1\textwidth]{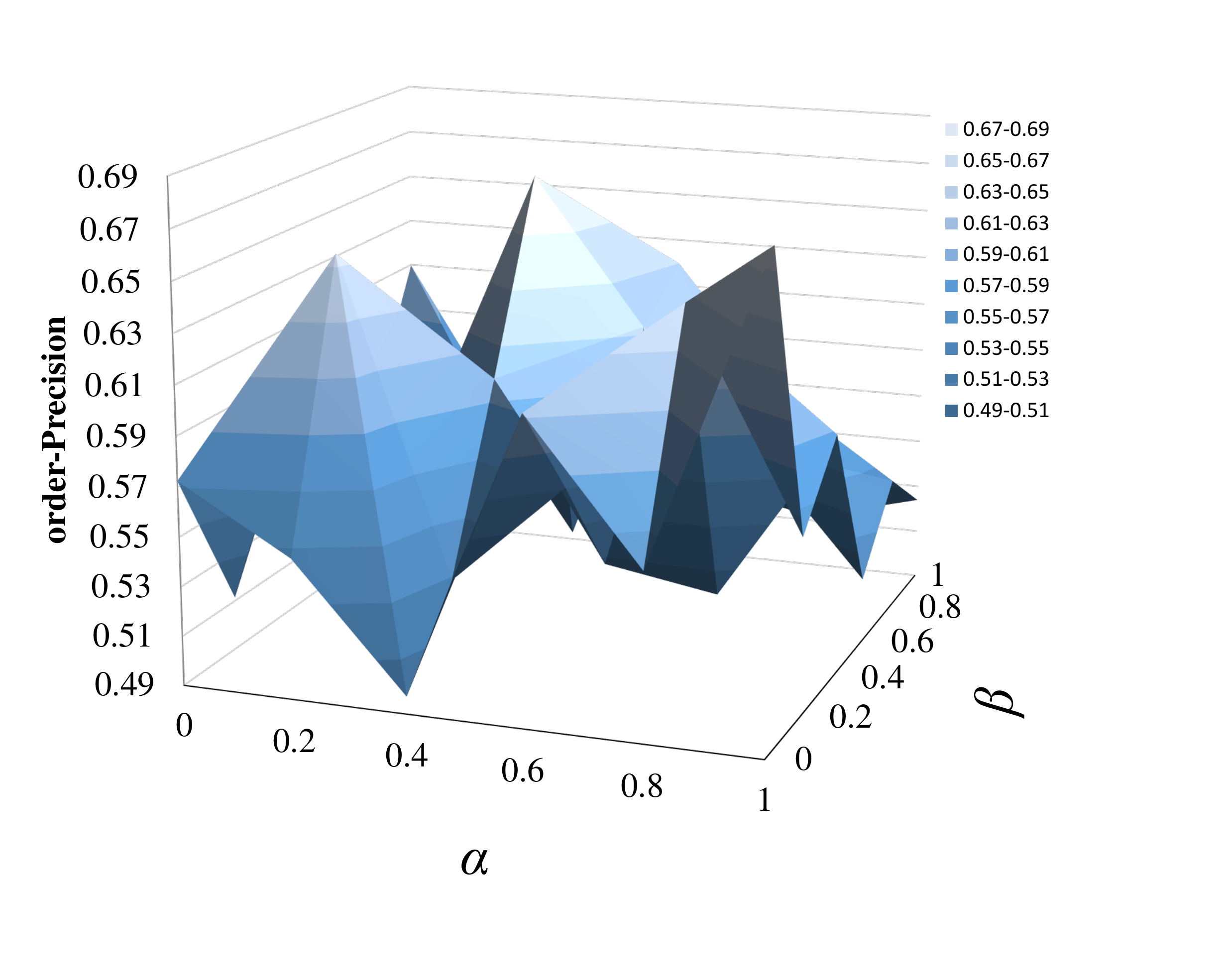}}
    \end{minipage}
    \caption{Tuning the parameter $\alpha$ and $\beta$ on Weibo.}
    \label{ParaInWeibo}
  \end{figure}

\subsection{Hyper-parameter Tuning}
In this subsection, we investigate the hyper-parameters $\alpha$ and $\beta$ in Equation (\ref{Eq_TL}) on the validation set, which control the influence of the cascading affinity and the structure proximity on the embedding learning, respectively.

For simplicity, we fix $\gamma = 0.002$ and adopt a grid search in the range of $[0, 1]$ with a step size of 0.2 to determine the optimal values of $\alpha$ and $\beta$. Fig. \ref{ParaInDigg}, Fig. \ref{ParaInTwitter}, and Fig. \ref{ParaInWeibo} show the results of MAP and order-Precision over different combinations of $\alpha$ and $\beta$ on three datasets. Through a comprehensive comparison, we can find that, in most cases the MAPs and order-Precisions at non-zero $\alpha$ and $\beta$ are better than those at zero $\alpha$ and $\beta$. Taking the Fig.\ref{ParaInDigg} (a) as an instance, the MAP value at (0.6, 0.8) is 0.8835, which is higher than 0.8703 at (0.0, 0.0). It verifies that appropriately applying cascading affinity and structural proximity as constrains can improve the learned embeddings for information cascade prediction. The combinations of $\alpha$ and $\beta$ at which the sum of MAP and order-Precision achieve the highest are chosen for the remaining experiments. Based on this criterion, we set ($\alpha$, $\beta$) as (0.1, 0.9) for Digg, (0.6, 0.8) for Twitter,  and (0.8, 0.2) for Weibo.

\subsection{Effectiveness}
In this section, we will analyze the experiments results in the tasks of infection prediction and infection order prediction, which are presented in Table \ref{Tab_MAP} and Figure \ref{OrderPres} respectively.

\subsubsection{Infection Prediction}
\begin{table}
\renewcommand\arraystretch{1.3}
\centering
\caption{MAP@$k$ on Digg, Twitter and Weibo datasets}
\begin{tabular}{|p{1.2cm}<{\centering}|c|p{1.2cm}<{\centering}|p{1.2cm}<{\centering}|p{1.2cm}<{\centering}|p{1.2cm}<{\centering}|p{1.2cm}<{\centering}|}
\hline
\multirow{2}{*}{Dataset}&\multirow{2}{*}{Method}&\multicolumn{5}{c|}{MAP@$k$ (\%)}\cr
\cline{3-7}
&&@100&@300&@500&@700&@900\cr
\cline{1-7}
\multirow{6}{*}{Digg}
&NetRate&1.108&5.749&10.933&16.618&24.043\cr
&CDK&27.951&39.766&52.032&65.220&80.408\cr
&Embedded-IC&2.084&9.073&23.314&47.249&78.066\cr
&Topo-LSTM&2.444&17.535&25.812&42.779&69.534\cr
&DCE-C&32.356&55.308&63.546&66.823&86.879\cr
&DCE&\textbf{47.497}&\textbf{72.952}&\textbf{76.694}&\textbf{84.250}&\textbf{91.362}\cr
\cline{1-7}
\multirow{6}{*}{Twitter}
&NetRate&0.140&2.550&6.724&15.058&30.572\cr
&CDK&9.512&22.724&34.701&48.162&63.315\cr
&Embedded-IC&0.751&4.740&12.568&24.985&43.347\cr
&Topo-LSTM&0.665&5.084&13.681&26.083&42.050\cr
&DCE-C&15.983&27.846&37.427&53.617&65.858\cr
&DCE&\textbf{16.376}&\textbf{29.773}&\textbf{40.690}&\textbf{56.301}&\textbf{69.863}\cr
\cline{1-7}
\multirow{6}{*}{Weibo}
&NetRate&0.469&2.696&7.724&15.280&25.583\cr
&CDK&1.124&11.510&25.348&41.810&54.429\cr
&Embedded-IC&0.185&3.988&9.706&18.965&30.738\cr
&Topo-LSTM&0.005&0.268&2.204&7.084&19.774\cr
&DCE-C&3.466&28.526&52.084&62.684&71.339\cr
&DCE&\textbf{10.506}&\textbf{30.986}&\textbf{53.555}&\textbf{64.533}&\textbf{72.746}\cr
\hline
\end{tabular}
\label{Tab_MAP}
\end{table}
Tables \ref{Tab_MAP} gives the MAPs of different methods for infection prediction task, with the best ones in each case being boldfaced. From Table \ref{Tab_MAP} we can make some analyses as follows:

\begin{enumerate}
\item The proposed DCE-C and DCE always outperform all baselines, giving improvements on the best baselines by $5.989\%$ (Twitter, MAP@500) to $33.186\%$ (Digg, MAP@300) relatively across all datasets. We can also find that DCE achieves better results than DCE-C in every case, and it proves that by using node collaborations as constrains, DCE can better characterize relations between nodes, which are important in information cascades.

\item The results show that, through collaboratively mapping the nodes into a latent space with a deep architecture, DCE can better capture deep and non-linear features of nodes in information cascades than Netrate, which estimates infection probability directly with a shallow probabilistic model.

\item In contrast with embedding baselines CDK, Embedded-IC, and Topo-LSTM, DCE's deep collaborative embedding architecture can better preserve the cascading characteristics and structural properties of nodes, which are crucial for infection prediction. Unlike CDK which assumes unrealistically that information diffusion is driven by the relations between source node and the others, in DCE all infected nodes are thought to have potential influence on the not yet infected ones and cascading context is employed to model their temporal relations. And as DCE makes no assumption about the underlying diffusion mechanism, it can better utilize the cascading contexts of nodes than Embedded-IC which is based on the IC model. Compared with Topo-LSTM that also adopts a deep model, DCE does not rely on the
    knowledge of the underlying diffusion network, which is usually difficult to obtain.

\end{enumerate}

\subsubsection{Infection Order Prediction}
  \begin{figure}[t]
    \begin{minipage}[t]{0.32\textwidth}
        \centering
         \subfigure[Digg]{\includegraphics[width=1\textwidth]{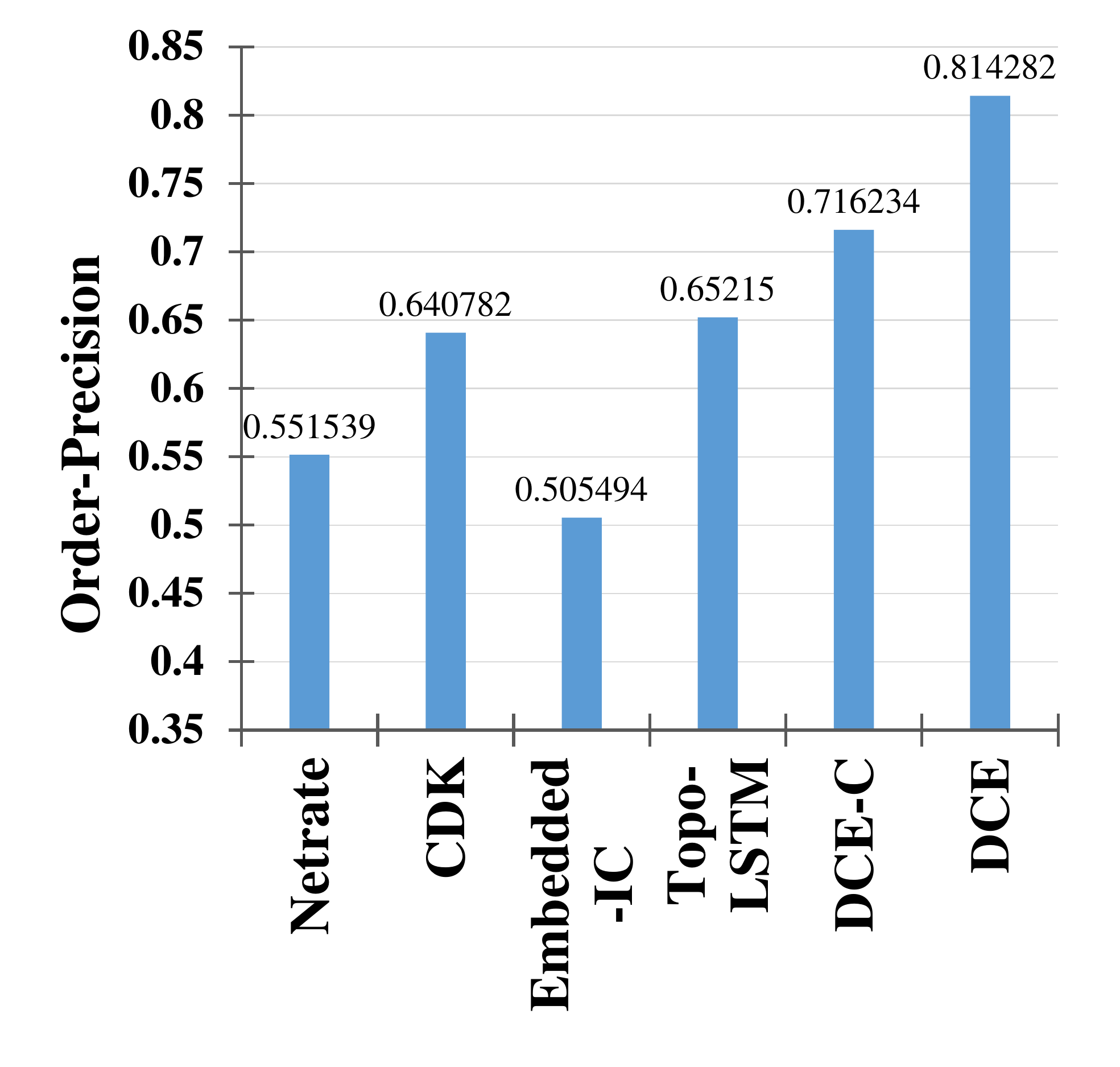}}
    \end{minipage}
    \begin{minipage}[t]{0.32\textwidth}
        \centering
         \subfigure[Twitter]{\includegraphics[width=1\textwidth]{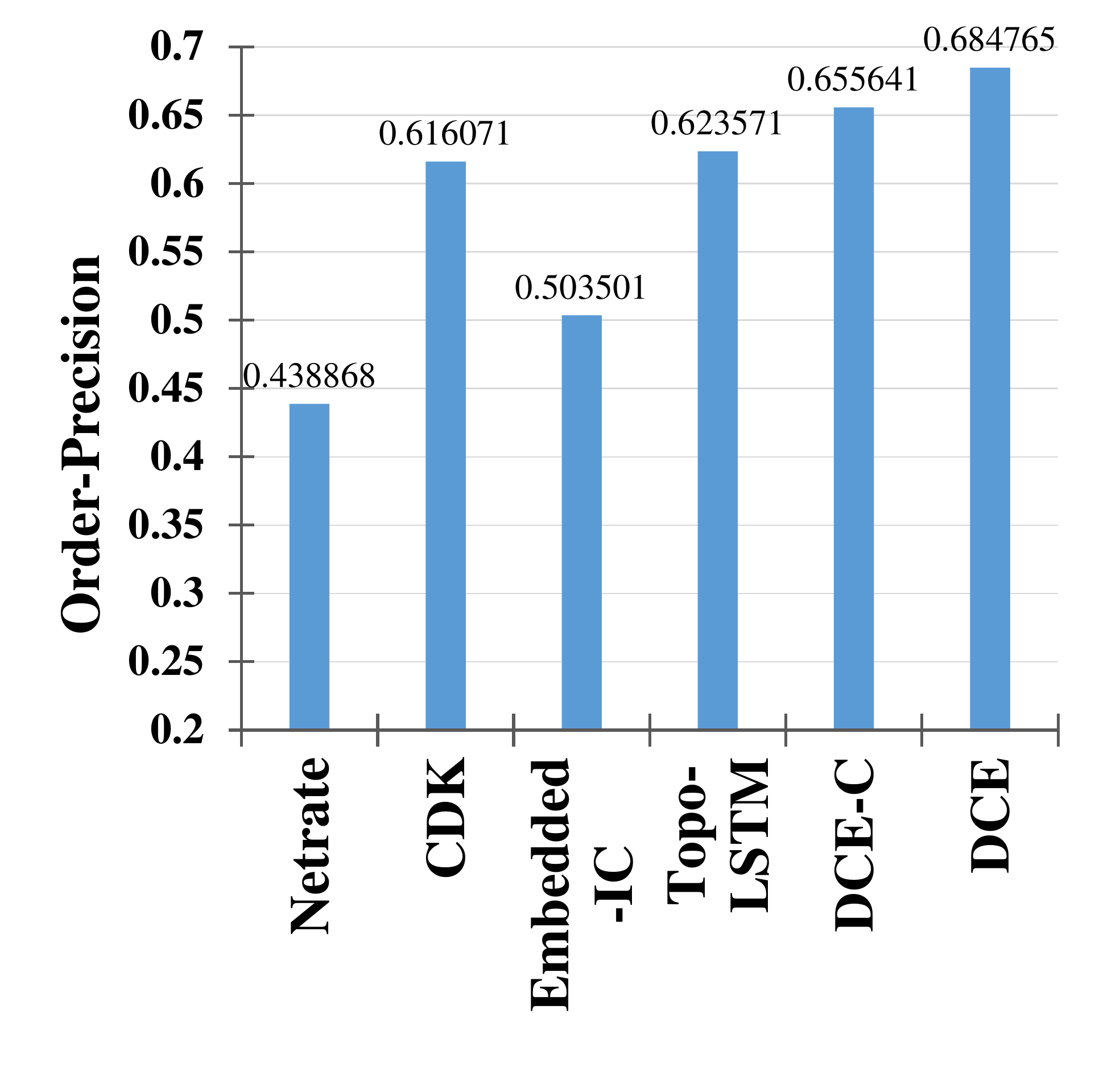}}
    \end{minipage}
    \begin{minipage}[t]{0.32\textwidth}
        \centering
         \subfigure[Weibo]{\includegraphics[width=1\textwidth]{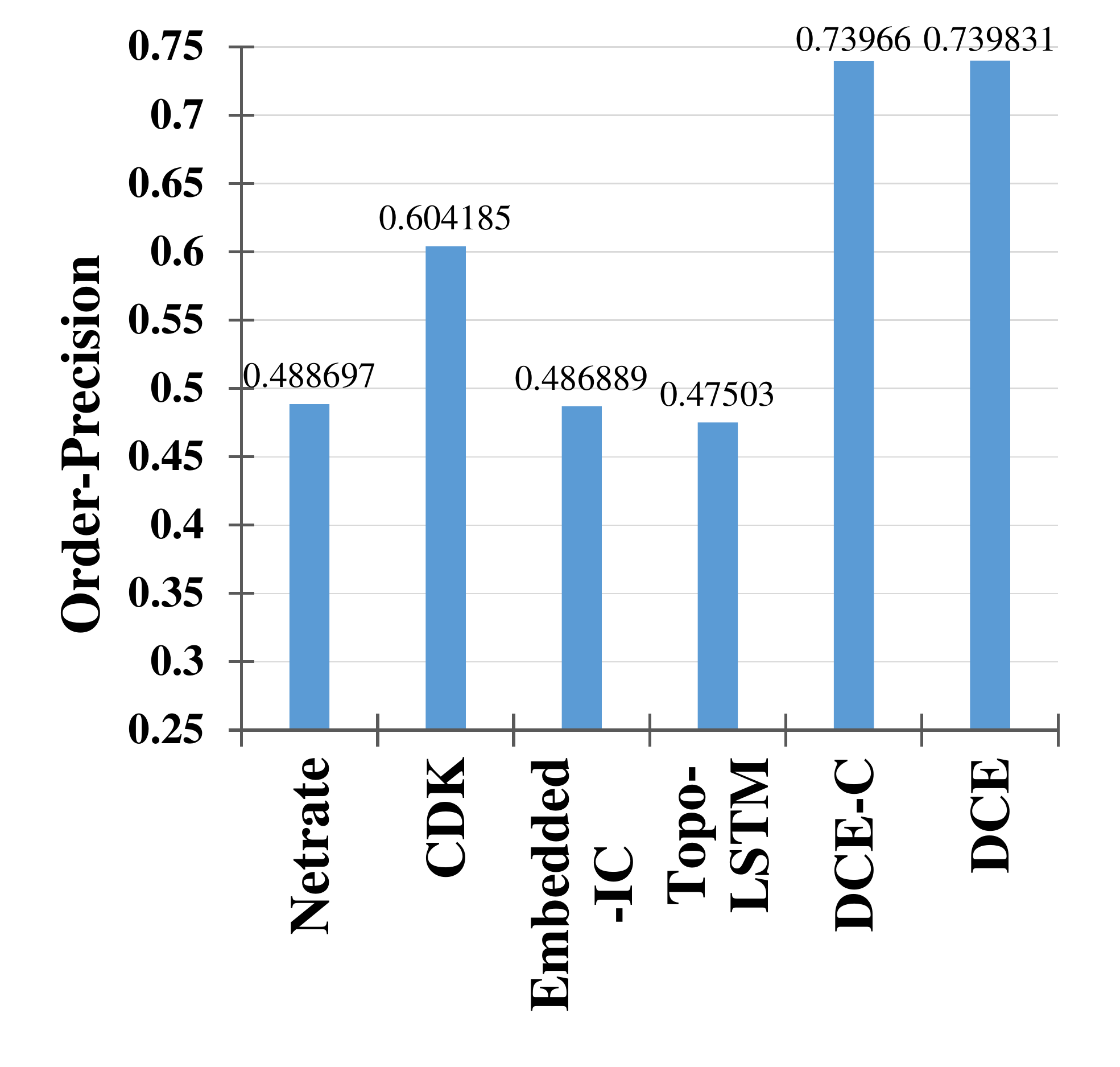}}
    \end{minipage}
    \caption{order-Precision on Digg, Twitter and Weibo datasets.}
    \label{OrderPres}
  \end{figure}

In Figure \ref{OrderPres} the order-Precisions of different methods for infection order prediction are presented, based on which several analyses can be made as follows:
\begin{enumerate}
\item We can see that the proposed DCE-C and DCE achieve best performance in all three datasets. The reason is that with the proposed cascading context, DCE is able to not only better preserve the temporal relations, but also better capture the infection order characteristics in information cascades than baselines. And DCE's superior results over DCE-C reveals that, even though cascading affinity and structural property do not indicate nodes infection orders explicitly, they can lead to further improvements when they are used as constrains in DCE.

\item To be more specific, NetRate is incapable of capturing the infection order features with its shallow probabilistic model. While CDK exploits heat diffusion kernel to formulate
    a ranking problem, where infection orders are kind of modeled, it can not fully characterize node infection order features like the proposed cascading context in DCE. For embedded-IC, nodes infection orders do not get any attention in this IC-based model and certainly can not be captured, which results in its bad performance. Notwithstanding Topo-LSTM's adoption of diffusion topology can encode the nodes infection orders to some extent, it still can not get rid of the dependence on the underlying diffusion network, which can not always be satisfied.

\end{enumerate}
\section {Related Work}
In this section, we briefly review two lines of related works with our research, including network embedding and information cascade prediction.

\subsection{Network Embedding}

With the wide employment of embedding methods in various machine learning tasks \cite{Mikolov2013Efficient,Mikolov2013Distributed,Marco2019Multilingual,Massimo2020Hybrid,Tingquan2020Low}, network embedding also gains more and more attentions and applications\cite{Cui2017A,HongYun2018AComprehensive}. Network embedding refers to assigning nodes in a network to low-dimensional representations and effectively preserving the network structure \cite{Cui2017A}. Intuitively, nodes can be represented by their correspondent raw or column feature vectors in the adjacent matrix of a network. However, sometimes these vectors are sparse with high dimensions, which brings challenges to machine learning tasks. As a result, a set of traditional network embedding methods \cite{Tenenbaum2000A,Roweis2000Nonlinear,Belkin2001Laplacian,Cox2001Multidimensional} are proposed mainly for dimension reduction. Nevertheless, these methods can only work well on networks of relatively small sizes and suffer from high computation cost when coping with online social networks with huge numbers of nodes.

Recent works like
DeepWalk \cite{Perozzi2014DeepWalk} and LINE \cite{Tang2015LINE} are proposed to learn low-dimensional representations for nodes through an optimization process instead of directly transforming the original feature vectors, where the scaling problem also can be well handled. Inspired by word2vec \cite{Mikolov2013Efficient,Mikolov2013Distributed}, DeepWalk considers the nodes in network as the words in natural language and utilizes random walks to generate node sequences, based on which the node representations are learned following the procedure of word2vec. As a more generalized version of DeepWalk, node2vec is proposed in \cite{Grover2016node2vec} with biased random walks to control the generation of nodes' contexts more flexibly. LINE produces embeddings for nodes with the expectation to preserve both the first-order and second-order proximities of the network neighborhood structure. Under the influence of these researches, a collection of network embedding methods are proposed for different scenarios. For instances, \cite{Swami2017metapath2vec} modifies DeepWalk for heterogeneous networks by introducing meta-path based random walks, and \cite{Xu2017Embedding} incorporates a harmonious embedding matrix to further embed the embeddings that only encode intra-network edges. As the deep neural network has shown remarkable effectiveness in many machine learning tasks, there also emerges a series of works which perform network embedding with a deep model. For example, \cite{Wang2016Structural} adopts a semi-supervised deep autoencoder model to exploit the first-order and second-order proximities jointly to preserve the network structure. \cite{Liao2017Attributed} learns nodes representations by keeping both the structural proximity and attribute proximity with a designed multilayered perceptron framework. And in \cite{Chang2015Heterogeneous}, the researchers use a highly nonlinear multilayered embedding function to capture the complex interactions between the heterogeneous data in a network.

However, most of these network embedding methods \cite{Yu2019TPNE,Palash2020dyngraph,Feiran2019Network} are not applicable to information cascade prediction. In our work, we employ an auto-encoder based collaborative embedding architecture to learn embeddings from nodes' cascading contexts with constrains.

\subsection{Information Cascade Prediction}
Information cascade phenomena have been widely investigated in the context of epidemiology, physical science and social science, and the development of online social network has greatly promoted related researches \cite{Chung2018Learning,Yuchen2018Influence,Devesh2017Predicting}. Most of the early researches \cite{Kempe2003Maximizing} analyse information cascade based on fixed models, the representatives among which are Independent Cascade(IC) \cite{Goldenberg2001Talk} model and the Linear Threshold(LT) \cite{Granovetter1978Threshold} model. Classic IC model treats the diffusion activity of information as cascades while the LT model determines infections of users according to thresholds of the influence pressure incoming from the neighborhood. Both of them can be unified into a same framework \cite{Kempe2003Maximizing}, and a series of extension work has been proposed \cite{Saito2008Prediction,Saito2010Generative,Saito2009Learning,Guille2012A,Wang2012Feature,Jingyi2019Influence,Furkan2018Influence}. For example, \cite{Saito2010Generative} extends the IC model to formulate a generative model that can take time delay into consideration. However, information diffusion processes are so complicated that we seldom exactly know
the underlying mechanisms of how information diffuses. What's more, these works are often based on the assumption that the explicit paths along which information propagates between nodes are observable, which is difficult to satisfy.

A collection of methods are proposed to infer the most possible links that can best explain the observed diffusion cascades without knowing the explicit paths. For instance, NetInf \cite{Gomez2011Inferring} and Connie \cite{Myers2010On} use greedy algorithms to find a fixed number of links between users that maximize the likelihood of a set of observed diffusions under an IC-like diffusion hypothesis. And a more general framework called NetRate \cite{Rodriguez2011Uncovering} has been proposed, which also occurs in our experiments as a baseline. NetRate models information diffusion process as discrete networks of continuous temporal process occurring at different rates, and then infers the edges of the global diffusion network and estimates the transmission rates of each edge that best explain the observed data \cite{Rodriguez2011Uncovering}. There are also further variants of this framework being proposed \cite{Gomez2013Modeling,Senzhang2014MMRate}. However, most of these works still rely on the assumption that information diffusion follows a parametric model.

In recent years, a set of researches \cite{Bourigault2014Learning,Gao2017A,Bourigault2016Representation,Wang2017Topological,Qiu2018DeepInf} which adopt network embedding techniques to handle information cascade prediction have be proposed. These methods usually embed nodes in a latent feature space, then the diffusion probabilities between nodes are computed based on their positions in the space. CDK proposed in \cite{Bourigault2014Learning} treats information diffusion as a ranking problem and maps nodes to a latent space using a heat diffusion process. However, it assumes the infected nodes orders of a cascade is influenced by the relations between source node and the other nodes, which is not realistic. \cite{Bourigault2016Representation} follows the mechanism of IC model to embed users in cascades into a latent space. \cite{Wang2017Topological} puts dynamic directed acyclic graphs into an LSTM-based model to generate topology-aware embeddings for nodes, which depends a lot on the network structure information. In contrast, our proposed method DCE collaboratively embed the nodes with a deep architecture into a latent space, without requirement of the knowledge about the underlying diffusion mechanisms and the explicit paths of diffusions on the network structure.

\section {Conclusions}
In this paper, we address the problem of information cascade prediction in online social networks with the network embedding techniques. We propose a novel model called Deep Collaborative Embedding (DCE) for information cascade prediction which can learn embeddings for not only infection prediction but also infection order prediction in a cascade, without the requirement to know the underlying diffusion mechanisms and the diffusion network. We propose an auto-encoder based collaborative embedding architecture to generate the embeddings that preserve the node structural property as well as the node cascading characteristics simultaneously in the learned embeddings. The results of extensive experiments conducted on real datasets verify the effectiveness of the proposed method.

\section*{Acknowledgment}
This work is supported by National Natural Science Foundation of China under grant 61972270, and in part by NSF under grants III-1526499, III-1763325, III-1909323, CNS-1930941, and CNS-1626432.

\section*{References}

\bibliographystyle{elsarticle-num}
\bibliography{DCECP}

\end{document}